\documentstyle[12pt]{article}

\begin{document}
\setcounter{page}{1}
\pagestyle{plain} \vspace{1cm}
\begin{center}
\Large{\bf Late-time acceleration and Phantom Divide Line Crossing with
Non-minimal Coupling and Lorentz Invariance Violation}\\
\small \vspace{1cm}
{\bf Kourosh Nozari}\quad\ and \quad {\bf S. Davood Sadatian}\\
\vspace{0.5cm} {\it Department of Physics,
Faculty of Basic Sciences,\\
University of Mazandaran,\\
P. O. Box 47416-1467,
Babolsar, IRAN\\
and\\
Research Institute for Astronomy and Astrophysics of Maragha, \\P.
O. Box 55134-441, Maragha, IRAN }\\

{\it knozari@umz.ac.ir\\
d.sadatian@umz.ac.ir}
\end{center}
\vspace{1.5cm}
\begin{abstract}
We consider two alternative dark energy models: a Lorentz invariance
preserving model with a nonminimally coupled scalar field and a
Lorentz invariance violating model with a minimally coupled scalar
field. We study accelerated expansion and dynamics of equation of
state parameter in these scenarios. While a minimally coupled scalar
field has not the capability to be a successful dark energy
candidate with cosmological constant line crossing, a nonminimally
coupled scalar field in the presence of Lorentz invariance or a
minimally coupled scalar field with Lorentz invariance violation
have this capability. In the later case, accelerated expansion and
phantom divide line crossing are the results of interactive nature
of this Lorentz violating scenario.\\
{\bf PACS}: 04.50.+h, 98.80.-k\\
{\bf Key Words}: Scaler-Vector-Tensor Theories, Braneworld
Cosmology, Accelerated Expansion, Lorentz Invariance Violation.
\end{abstract}
\vspace{1.5cm}
\newpage

\section{Introduction}
Recent evidences from supernova searches data [1,2], cosmic
microwave background (CMB) results [3-5] and also Wilkinson
Microwave Anisotropy Probe (WMAP) data [6,7], indicate an positively
accelerating phase of the cosmological expansion today and this
feature shows that the simple picture of universe consisting of
pressureless fluid is not enough. In this regard, the universe may
contain some sort of additional negative-pressure dark energy.
Analysis of the three year WMAP data [8-10] shows that there is no
indication for any significant deviations from Gaussianity and
adiabaticity of the CMB power spectrum and therefore suggests that
the universe is spatially flat to within the limits of observational
accuracy. Further, the combined analysis of the three-year WMAP data
with the supernova Legacy survey (SNLS) [8], constrains the equation
of state $w_{de}$, corresponding to almost ${74\%}$ contribution of
dark energy in the currently accelerating universe, to be very close
to that of the cosmological constant value. Moreover, observations
appear to favor a dark energy equation of state, $w_{de}<-1$ [11].
Therefore a viable cosmological model should admit a dynamical
equation of state that might have crossed the value $w_{de}= -1$, in
the recent epoch of cosmological evolution. In fact, to explain
positively accelerated expansion of the universe, there are two
alternative approaches: incorporating an additional cosmological
component or modifying gravity at cosmological scale.
Multi-component dark energy with at least one non-canonical phantom
field is a possible candidate of the first alternative. This
viewpoint has been studied extensively in literature ( see [12] and
references therein ). Another alternative to explain current
accelerated expansion of the universe is extension of general
relativity to more general theories on cosmological scales. In this
viewpoint, modified Einstein-Hilbert action resulting $f(R)$-gravity
( see [13] and references therein) and braneworld gravity [14-16]
are studied extensively. For instance, DGP (
Dvali-Gabadadze-Porrati) braneworld scenario as an IR modification
of general relativity explains accelerated expansion of the universe
in its positive branch via leakage of gravity to extra dimension. In
this model, equation of state parameter of dark energy never crosses
$\omega(z)=-1$ line, and universe eventually turns out to be de
Sitter phase. But, in this setup if we use a single scalar field
(ordinary or phantom) on the brane, we can show that equation of
state parameter of dark energy can cross phantom divide line (PDL)
[17]. Also quintessential behavior can be achieved in a geometrical
way in higher order theories of gravity [18].

From another view point, impact of Lorentz invariance violation
(LIV) on cosmology has been studied recently [19,20]. This issue has
been studied in the context of scalar-vector-tensor theories [19].
It has been shown that Lorentz violating vector fields affect the
dynamics of the inflationary models. One of the interesting feature
of this scenario is that the exact Lorentz violating inflationary
solutions are related to the absence of the inflaton potential. In
this case, the inflation is completely associated with the Lorentz
invarance violation and depends on the value of the coupling
parameters [20]. One important observation has been made in
reference [21] which accelerated expansion and crossing of phantom
divide line with one minimally coupled scalar field in the presence
of a Lorentz invariance violating vector field has been shown. This
model is essentially an interacting model which consists of
interaction between scalar field and Lorentz violating vector field.
One important consequence in quintessence model is the fact that a
single minimally coupled scalar field has not the capability to
explain crossing of phantom divide line, $\omega_{\phi}=-1$ [22].
However, a single but non-minimally coupled scalar field is enough
to cross the phantom divide line by its equation of state parameter
[12]. Currently, models of phantom divide line crossing are so
important that they can realize that which model is better than the
others to describe the nature of dark energy. In this respect,
possible crossing of phantom divide line by equation of state
parameter in model universes with a non-minimally coupled scalar
field and Lorentz invariance violation is important.

With this preliminaries, the purpose of this paper is to study late
time acceleration and phantom divide line crossing in two model
universes: a model universe with a nonminimally coupled scalar field
in the presence of Lorentz invariance symmetry and a model universe
with a Lorentz invariance violating dark energy component with
minimal coupling. In the former case we extend our study to moving
domain wall picture of braneworld scenario. In this regard, we first
study cosmological consequences of a non-minimally coupled scalar
field. In this stage, with a numerical analysis of parameters space
of the model, we show that accelerated expansion and crossing of
phantom divide line are explainable in the Jordan frame. By
transforming to Einstein's frame, we show that this model cannot
account for crossing of phantom divide line. Our strategy differs
with existing literature in its special kind of numerical reasoning
based on an appropriate ansatz. Then we extend this model to a
braneworld setup. In this extension, brane is considered to be a
moving domain wall in a background $5$-dimensional anti de
Sitter-Schwarzschild (AdSS$_{5}$) black hole bulk. In other words,
we consider a static bulk configuration with two $5$-dimensional
anti de Sitter-Schwarzschild black hole spaces joined by a moving
domain wall. Then we study dynamics of equation of state parameter
of a non-minimally coupled scalar field in this setup. This model
has also capability to explain accelerated expansion and phantom
divide line crossing in a fascinating manner. Then we summarize
cosmological equations of a Lorentz invariance violating model in
the spirit of Scalar-Vector-Tensor theories. We find a relation
between Lorentz Invariance violation parameter and dynamics of
scalar field. This relation explicitly shows the interactive nature
of our Lorentz invariance violating model. This model, with a
minimally coupled scalar field accounts for crossing of phantom
divide line and also accelerated expansion. There are three
important outcomes of our study: it is impossible to cross phantom
divide line with a single and minimally coupled scalar field but
non-minimal coupling of scalar field in the Jordan frame provides
such an important feature. A non-minimally coupled scalar field on
the moving domain wall is a good candidate for dark energy which
explains both late-time acceleration and phantom divide line
crossing. Also, a Lorentz invariance violation model with a
minimally coupled scalar field accounts for late-time acceleration
and phantom divide line crossing. It is important to note that
non-minimal coupling of a scalar field and gravity may provides a
basis for symmetry ( such as Lorentz invariance) breaking as has
been argued in reference [23]. To complete our study, based on
recent observational data we obtain some important constraints on
the parameters of the models in the favor of late-time accelerated
expansion.

\section{Accelerated Expansion and PDL Crossing with a Ricci-Coupled Scalar Field}
\subsection{The Jordan Frame}
For a model universe with a non-minimally coupled scalar field as
matter content of the universe, the action in the absence of other
matter sources in the Jordan frame can be written as follows
\begin{equation}
S=\int d^{4}x\sqrt{-g}\bigg[\frac{1}{{k_{4}}^{2}}\alpha(\phi)
R[g]-\frac{1}{2} g^{\mu\nu} \nabla_{\mu}\phi\nabla_{\nu}\phi
-V(\phi) \bigg],
\end{equation}
where we have included an explicit and general non-minimal coupling
of scalar field and gravity. For simplicity, from now on we set
${k_{4}}^{2}\equiv8\pi G_{N}=1$. Variation of the action with
respect to metric gives the Einstein equations
\begin{equation}
R_{\mu\nu}-\frac{1}{2}g_{\mu\nu}R=\alpha^{-1}{\cal{T}}_{\mu\nu}.
\end{equation}
${\cal{T}}_{\mu\nu}$, the energy-momentum tensor of the scalar field
non-minimally coupled to gravity, is given by
\begin{equation}
{\cal{T}}_{\mu\nu}=\nabla_{\mu}\phi\nabla_{\nu}\phi-\frac{1}{2}g_{\mu\nu}(\nabla\phi)^{2}-g_{\mu\nu}V(\phi)+
g_{\mu\nu}\Box\alpha(\phi)-\nabla_{\mu}\nabla_{\nu}\alpha(\phi),
\end{equation}
where $\Box$ shows 4-dimensional d'Alembertian. For FRW universe
with line element defined as
\begin{equation}
ds^{2}=-dt^{2}+a^{2}(t)d{\Sigma_{k}}^{2},
\end{equation}
where $d{\Sigma_{k}}^{2}$ is the line element for a manifold of
constant curvature $k = +1,0,-1$, the equation of motion for scalar
field $\phi$ is
\begin{equation}
\nabla^{\mu}\nabla_{\mu}\phi=V'-\alpha'R[g],
\end{equation}
where a prime denotes the derivative of any quantity with respect
to\, $\phi$. This equation can be rewritten as
\begin{equation}
\ddot{\phi}+3\frac{\dot{a}}{a}\dot{\phi}+\frac{dV}{d\phi}=
\alpha'R[g].
\end{equation}
where a dot denotes the derivative with respect to cosmic time\, $t$
\, and Ricci scalar is given by
\begin{equation}
R=6\bigg(\dot{H}+2H^{2}+\frac{k}{a^{2}}\bigg).
\end{equation}
With this non-minimally coupled scalar field as matter content of
the universe, cosmological dynamics are described by
\begin{equation}
\frac{\dot{a}^{2}}{a^{2}}=-\frac{k}{a^{2}}+\frac{\rho}{3},
\end{equation}
and
\begin{equation}
\frac{\ddot{a}}{a}=-\frac{1}{6}(\rho+3p).
\end{equation}
The effect of non-minimal coupling of scalar field and gravity is
hidden in the definition of $\rho$ and $p$. We assume that scalar
field, $\phi$, has only time dependence and using (3), we find
\begin{equation}
\rho=\alpha^{-1}\bigg(\frac{1}{2}\dot{\phi}^{2}+V(\phi)-6\alpha'H\dot{\phi}\bigg),
\end{equation}
\begin{equation}
p=\alpha^{-1}\bigg(\frac{1}{2}\dot{\phi}^{2}-V(\phi)+
2\Big(\alpha'\ddot{\phi}+2H\alpha'\dot{\phi}+\alpha''\dot{\phi}^2\Big)\bigg),
\end{equation}
where $H=\frac{\dot{a}}{a}$ is Hubble parameter. Now, equation (9)
takes the following form
\begin{equation}
\frac{\ddot{a}}{a}=-\frac{1}{6}\alpha^{-1}\bigg(2\dot{\phi}^{2}-2V(\phi)+
6\Big(\alpha'\ddot{\phi}+H\alpha'\dot{\phi}+\alpha''\dot{\phi}^2\Big)\bigg),
\end{equation}
and dynamics of equation of state parameter is given by
\begin{equation}
w\equiv\frac{p}{\rho}=\frac{\dot{\phi}^{2}-2V(\phi)+
4\Big(\alpha'\ddot{\phi}+2H\alpha'\dot{\phi}+\alpha''\dot{\phi}^2\Big)}
{\dot{\phi}^{2}+2V(\phi)-12\alpha'H\dot{\phi}}.
\end{equation}
From this equation, when $\dot{\phi}=0$, we obtain $p=-\rho$. In
this case $\rho$ is independent of $a$ and $V(\phi)$ plays the role
of a cosmological constant. In the minimal case when
$\dot{\phi}^{2}< V(\phi)$, using (9) we obtain $p<-\frac{\rho}{3}$
which shows an accelerated expansion which is driven by cosmological
constant. However, cosmological constant is not a good candidate for
dark energy since its suffers from several conceptual problems such
as its unknown origin and also need to huge amount of fine-tuning.
In non-minimal case the cosmological dynamics depends on the value
of non-minimal coupling. As a first goal in this paper we try to see
whether late-time accelerated expansion and crossing of phantom
divide line are explainable with this non-minimally coupled scalar
field as candidate for dark energy or not. Although this issue is
not new, our strategy for this purpose differs from existing
approaches ( see for instance [12]). To have positively accelerated
expansion we need $\rho+3p<0$ in equation (9). This is possible when
the following relation holds
\begin{equation}
\alpha^{-1}\Big[2\dot{\phi}^{2}-2V(\phi)+
6(\alpha'\ddot{\phi}+H\alpha'\dot{\phi}+\alpha''\dot{\phi}^2)\Big]<0.
\end{equation}
To proceed further, we assume a conformal coupling of scalar field
and gravity as $\alpha(\phi)=\frac{1}{2}\Big(1-\xi \phi^{2}\Big)$.
In fact in general relativity, and in all other metric theories of
gravity in which the scalar field is not part of the gravitational
sector, such a conformal coupling with $\xi=\frac{1}{6}$ is
necessary. Then we obtain
\begin{equation}
(1-3\xi)\dot{\phi}^{2}-V(\phi)+3\xi^{2}R\phi^{2}+6\xi
H\phi\dot{\phi}+3\xi\phi\frac{dV}{d\phi}<0.
\end{equation}
By imposing the weak energy condition and restricting study to the
case with $\xi\leq 1/6$, one finds [24]
\begin{equation}
-2V+3\xi\phi\frac{dV}{d\phi}<0
\end{equation}
and a necessary condition for cosmic acceleration is therefore
\begin{equation}
V-\frac{3\xi}{2}\phi\frac{dV}{d\phi}>0,\quad\quad\quad
\xi\leq\frac{1}{6}.
\end{equation}
\begin{figure}[htp]
\begin{center}\includegraphics{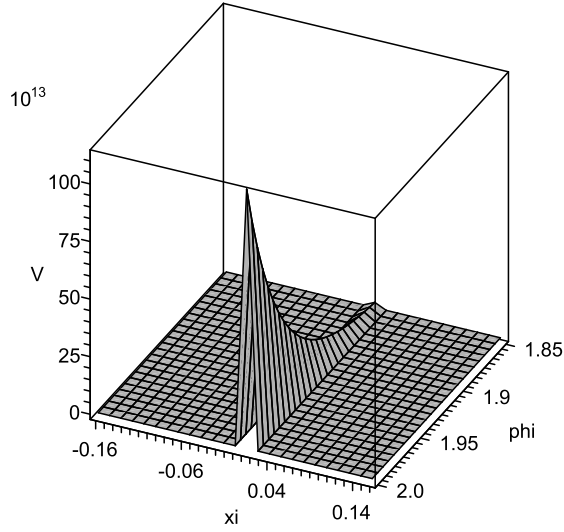} \vspace{6.5cm}
\end{center}
 \caption{\small {Variation of $V$
for different values of the $\phi$ and nonminimal coupling parameter
$\xi$ for $V = \phi^{\frac{2}{3\xi}}$.}}
\end{figure}
In this case to have cosmic acceleration with $\xi>0$, the potential
$V(\phi)$ should vary with $\phi$\, slower than power-law potential
$V_{c}(\phi)=V_{0}\Big(\frac{\phi}{\phi_{0}}\Big)^{\frac{2}{3\xi}}$.
However, when $\xi<0$, the necessary condition for cosmic
acceleration requires that $V$ grow faster than $V_{c}$ as $\phi$
increases [24]. As a specific example to show how this model works,
if we set $V(\phi)=\lambda \phi^{n}$,\, condition (17) gives
$\lambda\Big(1-\frac{3n\xi}{2}\Big)>0$\, which yields $\xi\leq
2/3n$. Figure $1$ gives a qualitative description of required
relation between potential and non-minimal coupling to have
accelerated expansion. As this figure shows, for positive $\xi$,
only for $0<\xi<0.026 $, this non-minimal model has the capability
to explain accelerated expansion. For a more general consideration,
we try a reliable ansatz so that
$\phi(t)\approx\frac{A}{t^{\beta}}$\, ( we assume a decreasing power
law ansatz for scalar field with $\beta>1$) and $a(t)\approx
Bt^{\nu}$. Accelerated expansion requires $\nu>1$. Now equation (12)
can be rewritten as follows
\begin{equation}
\nu(\nu-1)t^{-2}=-\frac{1}{3}\Big(1-\xi
t^{-2\beta}\Big)^{-1}\bigg\{2\beta^{2}t^{-2\beta-2}-2t^{-\beta
n}-6\Big[\beta(\beta+1)\xi t^{-2\beta-2}+\beta\nu \xi
t^{-2\beta-2}-\xi \beta^{2} t^{-2\beta-2}\Big]\bigg\}
\end{equation}
Considering terms of order $t^{-2\beta-2}$, we find
\begin{equation}
3\nu(\nu-1)\xi = 6\Big[\beta(\beta+1)\xi+\beta
\nu\xi-\xi\beta^{2}\Big]-2\beta^{2}.
\end{equation}
On the other hand, equation (6) for spatially flat FRW geometry
gives
\begin{equation}
\beta(\beta+1)t^{-\beta-2} -3\nu\beta t^{-\beta-2}+n t^{-\beta
n+\beta}=6\xi\nu t^{-\beta-2}-12\xi\nu^{2}t^{-\beta-2}.
\end{equation}
By considering terms of order $t^{-\beta-2}$, we find
\begin{equation}
\beta(\beta+1) -3\nu\beta=6\xi\nu-12\xi\nu^{2}.
\end{equation}
Now we have two equations (19) and (21) for three parameters $\xi$,
$\beta$ and $\nu$. We first solve equation (19) for $\nu$ to obtain
\begin{equation}
\nu=\frac{3\xi-6\beta\xi\pm
\sqrt{(3\xi-6\beta\xi)^{2}+12\xi\Big\{6\Big[\beta(\beta+1)\xi-\xi\beta^{2}\Big]-2\beta^{2}\Big\}}}{6\xi}.
\end{equation}
A numerical analysis shows that reality of $\nu$ is preserved if we
choose positive $\xi$ with $1\leq \beta<1.9$ ( note that this
condition is supported by equation (21) which gives
$\beta^{2}-2\beta+1\geq 0$). With this requirements and taking
positive sign in (22) we obtain possible values of $\nu$ in this
ansatz. The result is shown in figure $2$. This figure shows that
with nonminimally coupled scalar field one can explain accelerated
expansion, that is $\nu>1$, naturally. Thus non-minimal coupling of
scalar field and gravity in the Jordan frame provides a suitable
framework for explanation of late-time accelerated expansion. Note
that with negative sign in equation (22) it is impossible to find a
positive $\nu$.

\begin{figure}[htp]
\begin{center}\includegraphics{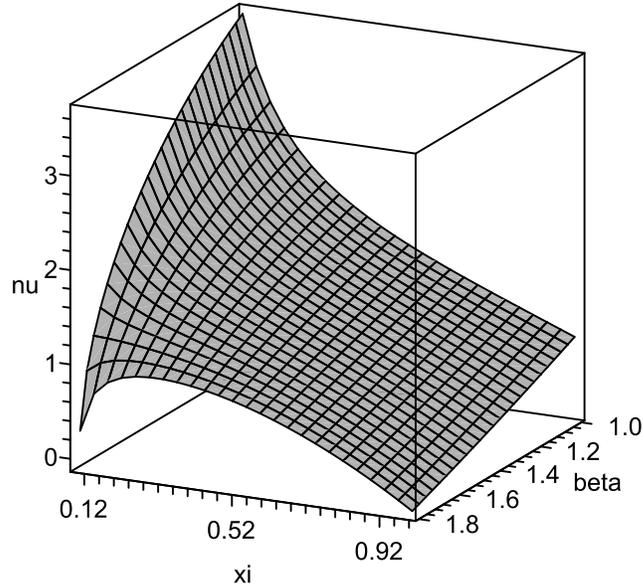} \vspace{7.5cm}
\end{center}
 \caption{\small {A non-minimally coupled scalar field in Jordan frame has the capability
 to explain late-time accelerated expansion with $\nu >1$ in parameter space.}}
\end{figure}

On the other hand, with above ansatz, dynamics of equation of state
parameter for a non-minimally coupled scalar field is given by
\begin{equation}
\omega=\frac{\beta^{2}t^{-2\beta-2}+2t^{-\beta
n}+4\Big[-\xi\beta(\beta+1)t^{-2\beta-2}+2\nu\xi\beta
t^{-2\beta-2}-\xi\beta^{2}t^{-2\beta-2}\Big]
}{\beta^{2}t^{-2\beta-2}+2t^{-\beta n}-12\beta\xi\nu t^{-2\beta-2}}.
\end{equation}
Figure $3$ shows the crossing of phantom divide line with equation
of state parameter of this non-minimally coupled scalar field. On
the other hand, as figure $4$ shows, in the case of $\xi=0$, that is
a single minimally coupled scalar field, there is no crossing of
phantom divide barrier, as has been emphasized by other literature
[22].
\begin{figure}[htp]
\begin{center}\includegraphics{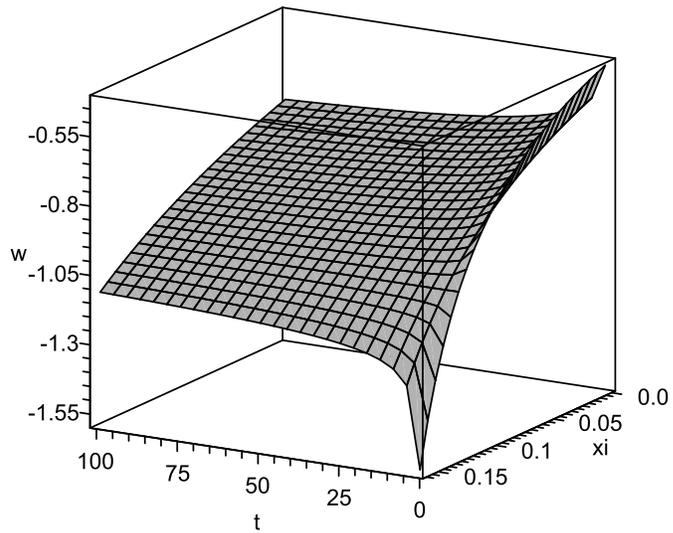} \vspace{5.5cm}
\end{center}
 \caption{\small {A non-minimally coupled scalar field in Jordan frame has the capability
 to have crossing of phantom divide line by its equation of state
 parameter in a suitable domain of parameter space. }}
\end{figure}

\begin{figure}[htp]
\begin{center}\includegraphics{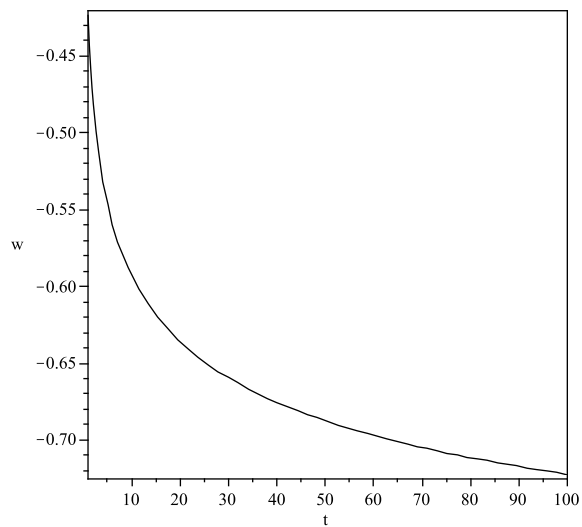} \vspace{5cm}
\end{center}
 \caption{\small {Equation of state parameter of a single minimally coupled scalar
 field ( with $\xi=0$ ), cannot explain crossing of phantom divide line [22].}}
\end{figure}
\subsection{The Einstein frame}
Now we study the situation in Einstein frame by a conformal
transformation. The action (1) in Jordan frame can be rewritten as
follows
\begin{eqnarray}
S=\int d^4x \sqrt{-g} \left[ \frac{1}{2}R
-\frac{1}{2}g^{\mu\nu}\partial_{\mu}\phi \partial_{\nu}\phi
-\frac{1}{2} \xi R \phi^2-V(\phi)\right]
\end{eqnarray}
where we assumed ${k_4}^{2}=1$ and
$\alpha(\phi)=\frac{1}{2}(1-\xi\phi^2)$ and $\xi$ is a non-minimal
coupling. The metric signature convention is chosen to be
$(+\,-\,-\,-)$ with spatially flat Robertson-Walker metric as
follows
\begin{eqnarray}
ds^2=dt^2-a^{2}(t)\delta_{ij}dx^idx^j.
\end{eqnarray}
To obtain the fundamental background equations in Einstein frame, we
perform the following conformal transformation
\begin{eqnarray}
\hat{g}_{\mu\nu}=\Omega g_{\mu\nu}, ~~~~~~ \Omega=1-\xi\phi^2.
\end{eqnarray}
Here we use a hat on a variable defined in the Einstein frame. The
conformal transformation gives
\begin{eqnarray}
S=\int d^4x \sqrt{-\hat{g}} \left[ \frac{1}{2}\hat{R}
-\frac{1}{2}F^2(\phi)\hat{g}^{\mu\nu}\partial_{\mu}\phi\partial_{\nu}\phi
-\hat{V}(\phi) \right],
\end{eqnarray}
where by definition
\begin{eqnarray}
F^2(\phi)\equiv\frac{1-\xi\phi^2(1-6\xi)}{(1-\xi\phi^2)^2}
\end{eqnarray}
and
\begin{eqnarray}
\hat{V}(\phi)\equiv\frac{V(\phi)}{(1-\xi\phi^2)^2}.
\end{eqnarray}
Therefore, one may redefine the scalar field as follows
\begin{eqnarray}
\frac{d\hat{\phi}}{d\phi}=F(\phi)=\frac{\sqrt{1-\xi\phi^2(1-6\xi)}}{1-\xi\phi^2}.
\end{eqnarray}
When we investigate the dynamics of universe in the Einstein frame,
we should transform our coordinates system to make the metric in the
Robertson-Walker form
\begin{eqnarray}
\hat{a}=\sqrt{\Omega}a,~~d\hat{t}=\sqrt{\Omega}dt,
\end{eqnarray}
and we obtain
\begin{eqnarray}
d\hat{s}^2=d\hat{t}^2-\hat{a}^2(\hat{t})\delta_{ij}dx^idx^j.
\end{eqnarray}
Note that the physical quantities in Einstein frame should be
defined in this coordinate system. Now the field equations can be
written as follows
\begin{eqnarray}
\hat{H}^2=\frac{1}{3}\left[\frac{1}{2}
\bigg(\frac{d\hat{\phi}}{d\hat{t}}\bigg)^2+\hat{V}(\hat{\phi})
\right]=\frac{\hat{\rho}}{3},
\end{eqnarray}
\begin{eqnarray}
\frac{d^2\hat{\phi}}{d\hat{t}^2}+3\hat{H}\frac{d\hat{\phi}}{d\hat{t}}+
\frac{d\hat{V}}{d\hat{\phi}}=0
\end{eqnarray}
where $\hat{H}=\frac{\hat{\dot{a}}}{\hat{a}}$. We assume that scalar
field $\hat{\phi}$ has only time dependence and we find dynamics of
equation of state as follows
\begin{equation}
\hat{\omega}_{\phi}=\frac{\hat{p}}{\hat{\rho}}=\frac{\frac{1}{2}
\bigg(\frac{d\hat{\phi}}{d\hat{t}}\bigg)^2-\hat{V}(\hat{\phi})}{\frac{1}{2}
\bigg(\frac{d\hat{\phi}}{d\hat{t}}\bigg)^2+\hat{V}(\hat{\phi})}.
\end{equation}
This is an interesting result: it shows that a non-minimally coupled
scalar field in Einstein frame cannot support the phantom phase. In
fact, conformal transformation from Jordan frame to Einstein frame
transforms the equation of state parameter to its minimal form but
with a redefined scalar field and in this case it is impossible to
achieve phantom phase ( and therefore no crossing of phantom divide
line).

\section{Braneworld Considerations}
In this section we show that a minimally coupled scalar field
localized on the brane provides even more suitable candidate for
explanation of accelerated expansion and phantom divide line
crossing. With this motivation, in which follows, along with studies
in [25-28], we consider a moving domain wall picture of braneworld
to discuss the issues of late-time acceleration and phantom divide
line crossing of equation of state parameter with a non-minimally
coupled scalar field localized on the brane. Following [26], we
consider a static bulk configuration with two $5$-dimensional anti
de Sitter-Schwarzschild (AdSS$_{5}$) black hole spaces joined by a
moving domain wall. To embed this moving domain wall into
$5$-dimensional bulk, it is then necessary to specify normal and
tangent vectors to this domain wall with careful determination of
normal direction to the brane. We assume that domain wall is located
at coordinate $r=a(\tau)$ where $a(\tau)$ is determined by Israel
junction conditions [29]. In this model, observers on the moving
domain wall interprets their motion through the static
$5$-dimensional bulk background as cosmological expansion or
contraction. Now consider the following line element [26]
\begin{equation}
{{dS}_{5\pm}}^{2}=-\bigg(k-\frac{\eta_{\pm}}{r^2}+\frac{r^2}{\ell^{2}}\bigg)dt^{2}
+\frac{1}{k-\frac{\eta_{\pm}}{r^2}+\frac{r^2}{\ell^{2}}}dr^{2}+r^{2}\gamma_{ij}dx^{i}dx^{j},
\end{equation}
where $\pm$ stands for left($-$) and right($+$) side of the moving
domain wall, $\ell$ is curvature radius of AdS$_{5}$ manifold and
$\gamma_{ij}$ is the horizon metric of a constant curvature manifold
with $k=-1,\, 0,\,1$ for open, flat and closed horizon geometry
respectively and $\eta_{\pm}\neq 0$ generates the electric part of
the Weyl tensor on each side. This line element shows a topological
anti de Sitter black hole geometry in each side. Using Israel
junction conditions [29] and Gauss-Codazzi equations we find the
following generalization of the Friedmann and acceleration equations
\begin{equation}
\frac{\dot{a}^{2}}{a^2}+\frac{k}{a^{2}}=\frac{\rho}{3}+\frac{\eta}{a^{4}}+
\frac{\ell^{2}}{36}\rho^{2},
\end{equation}
\begin{equation}
\frac{\ddot{a}}{a}=-\frac{\rho}{6}(1+3w)-\frac{\eta}{a^4}-\frac{\ell^2}{36}\rho^{2}(2+3w),
\end{equation}
where we have adapted a $Z_{2}$-symmetry with
$\eta_{+}=\eta_{-}\equiv\eta$ and $\omega$ is defined as
$\omega=\frac{p}{\rho}$. Assuming that brane is tensionless, in
which follows we discuss two cases with $\eta=0$  and  $\eta\neq
0$\, separately. Note that $\eta$ is the coefficient of a term which
is called holographic matter term. For $\eta=0$, each sub-manifolds
of bulk spacetime are exact AdS$_{5}$ spacetimes. Now we consider a
localized non-minimally coupled scalar field on the brane and
discuss its cosmological implications especially on late-time
dynamics. In this case we use energy density and pressure of scalar
field defined in equations (10) and (11) as the only matter source
on the brane. In this case, equation (38) takes the following form
$$\frac{\ddot{a}}{a}=-\frac{1}{6\alpha}\bigg(\frac{1}{2}\dot{\phi}^{2}+V(\phi)-
6\alpha'H\dot{\phi}\bigg)\bigg(1+3\frac{\dot{\phi}^{2}-2V(\phi)+
4\big(\alpha'\ddot{\phi}+2H\alpha'\dot{\phi}+\alpha''\dot{\phi}^2\big)}
{\dot{\phi}^{2}+2V(\phi)-12\alpha'H\dot{\phi}}
\bigg)$$
\begin{equation}
-\frac{\ell^2}{36\alpha^2}\bigg(\frac{1}{2}\dot{\phi}^{2}+V(\phi)-
6\alpha'H\dot{\phi}\bigg)^{2}\bigg(2+3\frac{\dot{\phi}^{2}-2V(\phi)+
4\big(\alpha'\ddot{\phi}+2H\alpha'\dot{\phi}+\alpha''\dot{\phi}^2\big)}
{\dot{\phi}^{2}+2V(\phi)-12\alpha'H\dot{\phi}}\bigg),
\end{equation}
where $H=\frac{\dot{a}}{a}$ is Hubble parameter on the moving domain
wall. This is a complicate relation and to explain its cosmological
implications, we have to consider either some limiting cases or
specify $\alpha(\phi)$, $V(\phi)$ and $\phi$. One can apply the
ansatz introduced in the last section with conformal coupling of
scalar field and Ricci scalar on the brane to investigate late time
behavior of this equation. But, due to existence of several
fine-tunable parameters and a combination of plus and minus signs in
this relation, essentially it is possible to find a domain of
parameters space that satisfies the condition $\ddot{a}>0$ in the
favor of positively accelerated expansion. For instance, if we set
$\phi=\phi_{0} e^{-\kappa t}$\, with $\kappa >0$, $a=a_{0} t^{\nu}$,
$V=\lambda \phi^{n}$, and  $A=\frac{\ddot{a}}{a}$, then equation
(39) gives
\begin{equation}
A=-{\frac {{\it E_{1}}\, \left( 1+3\,{\it E_{2}} \right)
}{3-3\,\xi\,\phi}}-{ \frac {{\ell}^{2}{{\it E_{1}}}^{2} \left(
2+3\,{\it E_{2}} \right) }{9-9\,\xi\, \phi}}
\end{equation}
where
$$E_{1}\equiv 0.5\,\kappa\,{e^{-2\,\kappa\,t}}+\lambda\,{e^{-\kappa\,nt}}-6\,{
\frac {\xi\,{e^{-2\,\kappa\,t}}\nu\,\kappa}{t}}$$ and
$$E_{2}\equiv \left(
{\kappa}^{2}{e^{-2\,\kappa\,t}}-2\,\lambda\,{e^{-\kappa\,nt}}+
8\,{\frac {\xi\,{e^{-2\,\kappa\,t}}\nu\,\kappa}{t}} \right)  \left(
{ \kappa}^{2}{e^{-2\,\kappa\,t}}+2\,\lambda\,{e^{-\kappa\,nt}}-12\,{
\frac {\xi\,{e^{-2\,\kappa\,t}}\nu\,\kappa}{t}} \right) ^{-1}.$$
Figure $5$ shows the possibility of accelerated expansion ( $A>0$
for $\nu>1$ in some appropriate domain of parameter space ( for
example with $\lambda=\kappa=\ell=1$ and $\xi=-0.1$)). The case with
$\eta\neq 0$ accounts for accelerated expansion in even more simpler
manner due to its wider parameter space. In this braneworld setup,
equation of state parameter with above ansatz ( defined before
equation (40)) has the following form
\begin{equation}
\omega(t)=\frac{{\kappa}^{2}{e^{-2\,\kappa\,t}}-2\,\lambda\,{e^{-\kappa\,nt}}+
8\,{\frac
{\nu\,\xi\,\kappa\,{e^{-2\,\kappa\,t}}}{t}}}{{\kappa}^{2}{e^{-2\,\kappa\,t}}+2\,\lambda\,{e^{-\kappa\,nt}}-12\,{
\frac {\nu\,\xi\,\kappa\,{e^{-2\,\kappa\,t}}}{t}}}
\end{equation}
Figure $6$ shows the dynamics of equation of state parameter in this
case with above mentioned ansatz. As this figure shows, equation of
state parameter crosses the phantom divide line $\omega=-1$. On the
other hand, crossing of phantom divide line with $\eta\neq0$ is
easily achieved due to wider parameter space in this case. As we
have emphasized in introduction, models of phantom divide line
crossing are so important that they can realize that which model is
better than the others to describe the nature of dark energy. In
this sense a non-minimally coupled scalar field on the brane
provides a good candidate for explaining accelerated expansion and
crossing of phantom divide line as a reliable candidate for dark
energy.
\begin{figure}[htp]
\begin{center}\includegraphics{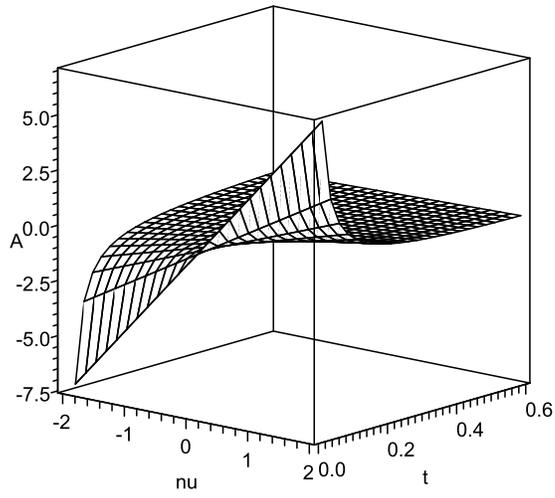} \vspace{4cm}
\end{center}
 \caption{\small {Accelerated expansion with a nonminimally coupled scalar
 field on the brane.}}
\end{figure}

\begin{figure}[htp]
\begin{center}\includegraphics{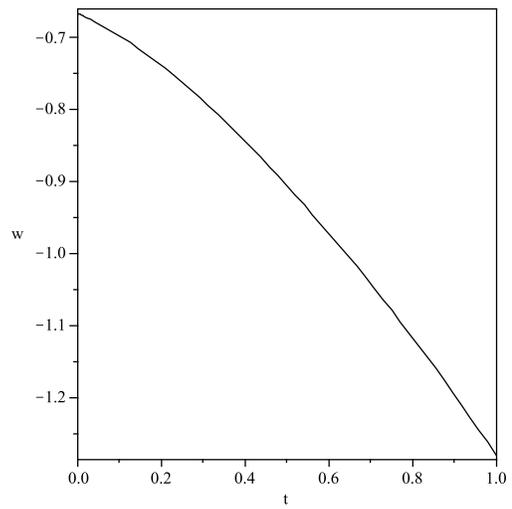} \vspace{3.5cm}
\end{center}
 \caption{\small {Crossing of phantom divide line with non-minimally coupled scalar field on the brane.}}
\end{figure}
\newpage
\section{Lorentz Invariance Violating Cosmology}
The purpose of this section is to study the effect of an explicit
violation of Lorentz invariance via incorporating a vector field in
the action. Following references [19,20], we summarize the
cosmological dynamics of a Lorentz invariance violating scenario
with a minimally coupled scalar field. Our goal is to find a
relation between Lorentz Invariance violation parameter and dynamics
of scalar field. This relation will affect the equation of state of
minimally coupled scalar field which is the central object of
subsequent discussions. In fact, this relation reflects the {\it
interactive} nature of this model. We start with the following
action for a typical scalar-vector-tensor theory which admits
Lorentz invariance violation
\begin{eqnarray}
     S= S_{g} + S_{u} + S_{\phi},
\end{eqnarray}
where the actions for the tensor field $S_g$, the vector field
$S_u$, and the scalar field $S_{\phi}$ are defined as follows
\begin{eqnarray}
        S_g &=& \int d^4 x \sqrt{-g}~ {1\over 16\pi G}R  \ ,
    \\
  S_u &=& \int d^4 x \sqrt{-g} \left[
  - \beta_1 \nabla^\mu u^\nu \nabla_\mu u_\nu
   -\beta_2 \nabla^\mu u^\nu \nabla_\nu u_\mu    -\beta_3 \left( \nabla_\mu u^\mu \right)^2 \right. \nonumber\\
 && \left.
  -\beta_4 u^\mu u^\nu \nabla_\mu u^\alpha \nabla_\nu u_\alpha
    + \lambda \left( u^\mu u_\mu +1 \right) \right]  \ ,
  \label{eq:act-VT} \\
  S_{\phi}  &=&  \int d^4 x \sqrt{-g}~ {\cal{L}}_{\phi}  \ .
   \end{eqnarray}
Action (42) is allowed to contain any non-gravitational degrees of
freedom in the framework of Lorentz violating scalar-tensor-vector
theory of gravity. As usual, we assume $u^\mu u_\mu = -1$ and that
the expectation value of vector field $u^\mu$ is $<0| u^\mu u_\mu
|0> = -1$\,\,[30]. $\beta_i(\phi)$ ($i=1,2,3,4$) are arbitrary
parameters with dimension of mass squared and ${\cal{L}}_{\phi}$ is
the Lagrangian density for scalar field. Also, $\sqrt{\beta_{i}}$
are mass scale of Lorentz symmetry breakdown\,[19,30]. The detailed
cosmological consequences of this action are studied in reference
[19]. Assuming a homogeneous and isotropic universe, we describe the
universe with the following metric
\begin{eqnarray}
ds^2 = - {\mathcal{N}}^2 (t) dt^2 + e^{2\alpha(t)} \delta_{ij} dx^i
dx^j \ ,
\end{eqnarray}
where ${\mathcal{N}}$ is a lapse function and the scale of the
universe is determined by $\alpha$. By variation of the action with
respect to metric and choosing a suitable gauge, one obtains the
following field equations
\begin{eqnarray}
   R_{\mu\nu}-{1\over 2}g_{\mu\nu}R = 8\pi G T_{\mu\nu} \ ,
   \end{eqnarray}
where $T_{\mu\nu} =T_{\mu\nu}^{(u)} + T_{\mu\nu}^{(\phi)}$ is the
total energy-momentum tensor, $T_{\mu\nu}^{(u)}$ and
$T_{\mu\nu}^{(\phi)}$ are the energy-momentum tensors of vector and
scalar fields, respectively. The time and space components of the
total energy-momentum tensor are given by [20]
\begin{eqnarray}
     T^{0}_{0} = - \rho_u -\rho_{\phi} \ , \qquad    T^{i}_i =  p_u+ p_{\phi} \ ,
     \end{eqnarray}
where the energy density and pressure of the vector field are
calculated as follows
\begin{eqnarray}
    && \rho_u =  -3\beta H^2  \ ,
   \\
    &&p_u =  \left(3 + 2{H^{\prime}\over H} + 2{\beta^{\prime}\over \beta} \right)\beta H^2 \ ,
     \\
    && \beta \equiv \beta_1 +3 \beta_2 + \beta_3 \ ,
    \end{eqnarray}
a prime denotes the derivative of any quantity with respect to
$\alpha$ \, and \, $H\equiv d\alpha/dt=\dot{\alpha}$ is the Hubble
parameter. One can see that $\beta_4$ does not contribute to the
background dynamics [19,20]. The energy equations for the vector
field $u$ and scalar field, $\phi$ are as follows
\begin{eqnarray}
   {\rho}^{\prime}_u + 3({\rho}_u + p_u)=+3H^2 \beta^{\prime}  \ ,
   \end{eqnarray}
\begin{eqnarray}
    {\rho}^{\prime}_{\phi} + 3({\rho}_{\phi} + p_{\phi})=-3H^2 \beta^{\prime}  \
    ,
   \end{eqnarray}
respectively. There is a non-conservation scheme in this setup due
to energy-momentum transfer between scalar and vector fields. This
is very similar to the case studied by Zimdahl {\it et al} [35]. As
they have shown, a coupling between a quintessence scalar field and
a cold dark matter (CDM) fluid leads to a stable, constant ratio for
the energy densities of both component compatible with a power law
accelerated cosmic expansion. In fact this coupling is responsible
for accelerated expansion and possible crossing of PDL line. In our
Lorentz invariance violating scenario this coupling is present
between scalar field and vector field as is manifested from
equations (52) and (53) corresponding to equations (4) and (5) of
Ref. [35] with $\delta\equiv
-3H^{2}\dot{\beta}\Pi_{u}=3H^{2}\dot{\beta}\Pi_{s}$, where $\Pi_{u}$
and $\Pi_{u}$ are effective pressure of vector and scalar component.
Nevertheless, the total energy equation in the presence of both the
vector and the scalar fields reads
\begin{equation}
   {\rho}^{\prime} + 3({\rho} + p)=0 \ , \quad (\rho = \rho_u +
   \rho_{\phi}),
   \end{equation}
which shows the conservation of total energy density. With these
preliminaries, dynamics of the model is described by the following
Friedmann equations
\begin{eqnarray}
     \left( 1 + \frac{1}{8\pi G \beta} \right) H^2={1\over 3\beta} \rho_{\phi}   \ ,
    \\
     \left( 1 + \frac{1}{8\pi G \beta} \right) \left( HH'+H^2\right)=-{1\over 6} \left( {\rho_{\phi}\over \beta} + {3p_{\phi}\over \beta} \right) - H^2 {\beta'\over \beta} \ .
    \end{eqnarray}
In the absence of vector field, that is, when  all \,$\beta_i =0$,
one recovers the standard equations of dynamics. For the scalar
sector of our model we assume the following Lagrangian
\begin{eqnarray}
{\cal{L}}_{\phi}= -{\varepsilon\over 2}(\nabla \phi)^2 - V(\phi) \ ,
\end{eqnarray}
where $(\nabla
\phi)^2=g^{\mu\nu}\partial_{\mu}\phi\partial_{\nu}\phi$. Ordinary
scalar fields are correspond to $\varepsilon = 1$ while $\varepsilon
= -1$ describes phantom fields. For the homogeneous scalar field,
the density $\rho_{\phi}$ and pressure $p_{\phi}$ are given as
follows
\begin{eqnarray}
&&\rho_{\phi} = {\varepsilon\over 2} H^2 \phi^{\prime 2} + V(\phi) \
,
\\
&&p_{\phi}= {\varepsilon\over 2} H^2 \phi^{\prime 2} - V(\phi) \ .
\end{eqnarray}
The corresponding equation of state parameter is
\begin{eqnarray}
\omega_{\phi}={p_{\phi}\over \rho_{\phi}} = - \frac{1- \varepsilon
H^2 \phi^{\prime 2}/2V}{1 + \varepsilon H^2 \phi^{\prime 2}/2V} \ .
\end{eqnarray}
Now the Friedmann equation takes the following form \,[20]
\begin{eqnarray}
H^2  = \frac{1}{3\bar{\beta}} \left[
  \frac{\varepsilon}{2} H^2 \phi^{\prime 2} + V(\phi) \right] ,
  \end{eqnarray}
where $\bar{\beta}=\beta+\frac{1}{8\pi G}$. Using this equation we
can show that
\begin{eqnarray}
  \phi^{\prime}=-2\varepsilon \bar{\beta}\left(\frac{H_{,\phi}}{H} +
\frac{\bar{\beta}_{,\phi}}{\bar{\beta}} \right) \ .
     \end{eqnarray}
Substituting this equation into the Friedmann equation, the
potential of the scalar field can be written as
\begin{eqnarray}
    V = 3\bar{\beta} H^2 \left[ 1-{2\over 3}
    \varepsilon\bar{\beta}\left({\bar{\beta}_{,\phi}\over \bar{\beta}}
    + {H_{,\phi}\over H}\right)^2 \right] \ .
\end{eqnarray}
Note that in the above equations the Hubble parameter $H$ has been
expressed as a function of $\phi$, $H=H(\phi(t))$. One can show that
the equation of state has the following form
\begin{eqnarray}
    \omega_\phi &=& -1 + {4\over 3}\varepsilon\bar{\beta}\left(\frac{H_{,\phi}}{H} +
    \frac{\bar{\beta}_{,\phi}}{\bar{\beta}} \right)^2 \nonumber\\
    &=&-1 + {1\over 3}\varepsilon \frac{\phi^{\prime 2}}{\bar{\beta}} \ .
    \end{eqnarray}
Equations (62) and (64) are essential equations in forthcoming
arguments. We stress here that violation of the Lorentz invariance
which has been introduced by existence of a vector field in the
action, now has incorporated in the dynamics of scalar field and
equation of state via existence of $\bar{\beta}$ reflecting
interactive nature of the model. This interesting feature allows us
to study crossing of phantom divide line by equation of state
parameter of minimally coupled scalar field and late-time
accelerated expansion as a result of interaction in the context of
Lorentz invariance violation. We need to solve these two equations,
(62) and (64),\, to find dynamics of scalar field $\phi$ and the
equation of state $\omega_\phi$. This will be achieved only if the
Hubble parameter $H(\phi)$ and the vector field coupling,\,
${\bar{\beta}}(\phi)$ are known. In which follows, our strategy is
to choose some different cases of the Hubble parameter $H(\phi)$ and
the vector field coupling ${\bar{\beta}}(\phi)$ and then
investigating possible crossing of phantom divide barrier and
late-time acceleration in this context. We obtain suitable domains
of parameter space which have the capability to explain late-time
acceleration and crossing of phantom divide line by equation of
state parameter.

\subsection{Late-Time Acceleration}
In reference [21] we have studied late-time acceleration and phantom
divide line crossing with Lorentz invariance violating fields for
several interesting cases. Here we extend that study for a more
general case. The condition for positive acceleration of the
universe, that is, $\ddot{a}>0$ can be rewritten as  $H'/H > -1$ in
this Lorentz invariance violating scenario. We consider a general
case where both the vector field coupling and the Hubble parameter
are functions of scalar field $\phi$ defined as follows ( see [21]
for motivations behind choosing this ansatz)
\begin{equation}
    H=H_0\phi^{\zeta} \ , \quad  \bar{\beta}(\phi) = m\phi^n \ ,\quad n >2 \
        \end{equation}
In which follows, we consider just a quintessence scalar field with
$\varepsilon = 1$. Using equation (62), for this case we obtain
\begin{equation}
\phi \left( t \right) = \Big[H_0(t-t_0)(-4\,\zeta m+4\,\zeta
mn+2\,{\zeta}^{2}m-4 \,mn+2\,m{n}^{2})+\phi_0 \Big ] ^{-
\left(\frac{1}{ n+\zeta-2} \right) }
\end{equation}
and using equation (64) we find
\begin{equation}
\omega_\phi(t)=-1+\frac{4}{3}m\phi^{n-2}(t)(\zeta+n)^2
\end{equation}
Now we obtain a condition for positively accelerated expansion,\,
$H'/H > -1$.\, We use equation (64), (65) and (67) to find
\begin{equation}
m^2<\frac{1}{4(-1)^n\phi^{n-2}(t)(\zeta+n)^2},~~~~~~ n>2 ,
\end{equation}
This relation can be used to constraint parameters of this model in
order to have late-time acceleration by confrontation with
observational data. Now we use relations (65) and (66) to obtain
dynamics of scale factor
\begin{equation}
a(t)=a_0(t_0)e^{\Upsilon}
\end{equation}
where
$$\Upsilon\equiv{\frac{2\phi_0^{\frac{\zeta}{(n+\zeta-2)}}
\Big[m(n+\zeta)(n+\zeta-2)H_0(t-t_0)+\frac{1}{2}\phi_0\Big]e^{\Omega}-\phi_0}{2\phi_0^{\frac{\zeta}{(n+\zeta-2)}}(n-2)m(n+\zeta)}}$$
and
$$\Omega\equiv-\zeta\bigg(\frac{ln\Big[2m(n+\zeta)(n+\zeta-2)H_0(t-t_0)+\phi_0\Big]}
{n+\zeta-2}\bigg)$$ The functional form of scale factor in this case
is very complicate. To find an understandable relation, we expand
relation (69) in Taylor series. Choosing $n=3$ , $\zeta=-2$ and
$m=-0.1$, we find
\begin{equation}
a(t)=0.286504+0.447663\,\,t+0.237821\,\,t^2+0.0352651\,\,t^3+O(t^4)
\end{equation}
This relation shows that a Lorentz invariance dark energy model
explicitly accounts for cosmic accelerated expansion as a result of
interactive nature of the model. Figure $7$ shows the variation of
scale factor with time. Evidently, it has positive second derivation
and so accounts for accelerated expansion.
\begin{figure}[htp]
\begin{center}\includegraphics{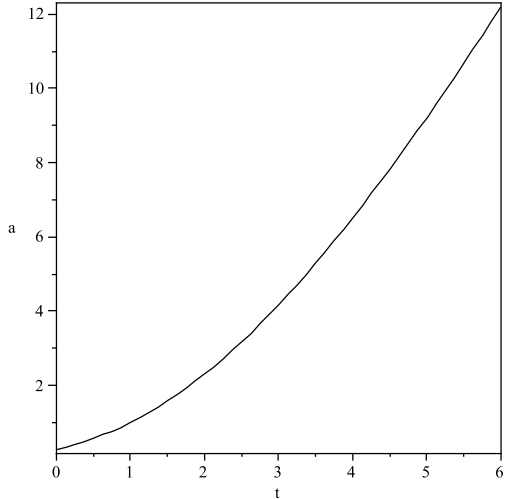} \vspace{6cm}
\end{center}
 \caption{\small {Variation of scale factor $a(t)$
for different values of $t$ with  $n=3$ , $\zeta=-2$ and $m=-0.1$.
This figure shows accelerated expansion in this Lorentz invariance
violating setup. The values of $\zeta$ are determined by relation
(56). }}
\end{figure}

\subsection{Crossing the Phantom Divide Line}
We can obtain dynamics of equation of state parameter for minimally
coupled scalar field in this Lorentz invariance violating model.
With $\phi$ defined as (66), the equation of state takes the
following form
\begin{equation}
\omega_\phi(t)=-1+\frac{4}{3}m\frac{(\zeta+n)^2}{\Bigg[H_0(t-t_0)(-4\,\zeta
m+4\,\zeta mn+2\,{\zeta}^{2}m-4 \,mn+2\,m{n}^{2})+\phi_0 \Bigg] ^{
\left(\frac{n-2}{ n+\zeta-2} \right) }},
\end{equation}
which explicitly has a dynamical behavior. This model allows us to
choose a suitable parameter space to explain crossing of phantom
divide barrier by equation of state parameter. This parameter space
should be checked by observational data in order to have a
reasonable cosmological model.

As an important point, we should be careful to choose the
appropriate equation of state for components that are used to
describe the universe energy-momentum content. As we have emphasized
earlier, a suitable coupling between a quintessence scalar field and
other matter content can leads to a constant ratio of the energy
densities of both components which is compatible with an accelerated
expansion of the Universe or Crossing of phantom divide line (for
more details see [35] and reference therein). In this respect and
for instance, the holographic dark energy models studied in Ref.
[22] have the phantom phase by adopting a native equation of state,
whereas the authors in [36] have found accelerating phase only using
the effective equation of state.\, Based on these arguments, we
should explain what kind of equation of state is used for observing
the nature of mixed fluids here. In our model, we have three sources
of energy-momentum: 1- standard ordinary matter, 2- Scalar Field as
a candidate of Dark Energy and 3- energy-momentum content depended
on Lorentz violating vector field. Here we assume that standard
matter has negligible contribution on the total energy-momentum
content of the universe and we can consider a constant linear
isothermal equation of state as $p_m=(\gamma-1)\rho_m$ that
$1\leq\gamma\leq 2$ for it. For other two energy-momentum contents,
it is possible to use the "trigger mechanism" to explain dynamical
equation of state. This means that we assume scalar- vector-tensor
theory containing Lorentz invariance violation which acts like the
hybrid inflation models. In this situation, vector and scaler field
play the roles of inflaton and the "waterfall" field respectively.
In this regard, we can fine-tune parameter $m$ in equation (68) to
obtain best fit model using the observational data. Of course, an
attractor solutions and fine-tuning in Lorentz violation model for
suitable inflation phase has been studied in Ref.[20]. Therefore it
is reasonable to expect that one of them will eventually dominate to
explain inflation or accelerating phase and crossing of phantom
divided line.

Figure $8$ shows the dynamics of $\omega(t)$ in this setup, it
crosses phantom divide line explicitly.  The most important aspect
of the present model is the fact that, Lorentz invariance violation
provides a situation that one scalar field and another vector field,
in an interactive picture, describe the phantom divide line crossing
and universe late time acceleration.
\begin{figure}[htp]
\begin{center}\includegraphics{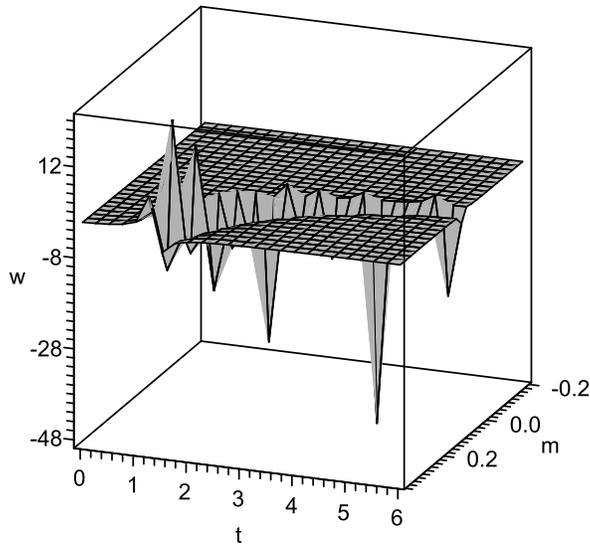} \vspace{7cm}
\end{center}
 \caption{\small {Variation of $\omega_\phi$
for different values of the vector field coupling $m$ and $t$ for
$n=3$ and $\zeta=-2$. Positive values of $\zeta$ show no phantom
divided barrier crossing. The values of $\zeta$  are determined by
relation (56). }}
\end{figure}
Figure $8$ may be used also to explain why we are living in an epoch
of \, $\omega< -1$ since in late time we see that $\omega< -1$. This
is the second cosmological coincidence problem. Remember that
$\bar{\beta}(t)$ plays the role of Lorentz invariance violation in
this setup. Equation of dynamics for $\bar{\beta}(t)$ implicitly has
an important meaning: by a suitable fine tuning one can construct a
Lorentz violating cosmology consistent with observational data. In
another words, this setup provides an important basis for testing
LIV in cosmological context. Although many different models can also
lead to phantom divided barrier crossing, our model is special in
this respect since it contains only one scalar field and the
presence of Lorentz violating vector field and interactive nature of
this model control the crossing. In this sense, fine tuning of
parameters space based on observational data restricts the value
that $\bar{\beta}(t)$ can attain. Any non-vanishing value of
$\bar{\beta}$ in our model shows violation of Lorentz symmetry in
this cosmological setup.  Lorentz invariance violating inflation
models constraint by WMAP and other observational data may provide
other tests of LIV in cosmological setup. To see possible detection
of Lorentz-violating fields in cosmology see [31,32,33]. In the
presence of LIV, just one scalar field is enough to achieve phantom
divide barrier crossing and existence of vector field controls the
situation. Two important point should be stressed here: firstly, as
figure $8$ shows, there are some sudden jumps of the equation of
state. In many existing models whose equation of state can cross the
phantom divide line, $\omega$ undulates around $-1$ randomly ([34]
and references therein). These jumps are actually a manifestation of
this undulation which may be a signature of chaotic behavior of
equation of state during its evolution. Secondly, as these figures
show, crossing of the phantom divide line can occur at late-time.
This fact, as second cosmological coincidence, needs additional
fine-tuning in model parameters and trigger mechanism, for instance,
can be used to alleviate this coincidence.

\section{Summary}
Light-curves analysis of several hundreds type Ia supernovae, WMAP
observations of the cosmic microwave background radiation and other
CMB-based experiments have shown that our universe is currently in a
period of accelerated expansion. In this respect, construction of
theoretical frameworks with potential to describe positively
accelerated expansion and crossing of the phantom divide line by
equation of state parameter, itself is an interesting challenge.
According to existing literature on dark energy models, a {\it
minimally} coupled scalar field is not a good candidate for dark
energy model with equation of state parameter crossing the phantom
divide line. On the other hand, a scalar field {\it non-minimally}
coupled to gravity in the Jordan frame has the capability to be a
suitable candidate for dark energy which provides this facilities.
Although this issue has been studied in literature, our study here
is different in its different approach based on numerical analysis
of parameter space. We have extended this study to a barneworld
setup where brane has been considered as a moving domain wall in a
static bulk background. In this braneworld setup, non-minimally
coupled scalar field provides even more reliable candidate for dark
energy. Then we have extended our study to the Lorentz invariance
violating dark energy model. We have shown that a minimally coupled
scalar field in the presence of a Lorentz violating vector field
provides a good candidate for dark energy with capability of
describing late-time acceleration and phantom divide line crossing.
One important observation here is the fact that this model achieve
an interactive nature which this interaction is responsible for
late-time acceleration and phantom divide line crossing. As some
details of our analysis, we emphasize that due to complication of
dynamical equations, we have restricted our study to some specific
form of non-minimal coupling and scalar field potentials and also we
have considered some special form of time evolution for scale factor
and scalar field. These choices, though especial, are natural and
motivated from powerful grounds based on recent observational data.
Crossing of phantom divide barrier by a single scalar field in the
presence of a Lorentz violating vector field and with suitable fine
tuning of model parameters in an interactive picture, is an
important outcome in this context. This feature is more considerable
were we emphasize that in the absence of Lorentz invariance
violating vector field, it is impossible to cross phantom divide
line just by one scalar field minimally coupled to gravity.\\

{\bf Acknowledgment}\\
This work has been supported partially by Research Institute for
Astronomy and Astrophysics of Maragha, Iran. We would like to thank
an anonymous referee for his/her important contribution in this
work.

\end{document}